\documentclass[aps,twocolumn,superscriptaddress,showpacs]{revtex4}%
\usepackage{amsfonts}
\usepackage{amsmath}
\usepackage{amssymb}
\usepackage{graphicx}%
\providecommand{\U}[1]{\protect\rule{.1in}{.1in}}
\providecommand{\U}[1]{\protect\rule{.1in}{.1in}}

\begin{document}
\title{optomechanically induced amplification and perfect transparency in double-cavity optomechanics}
\author{\ Xiao-Bo Yan}
\email{xiaoboyan@126.com}
\affiliation{College of Electronic Science, Northeast Petroleum University, Daqing, 163318,
P. R. China}
\affiliation{College of Physics, Jilin University, Changchun 130012, P. R. China}
\author{\ W. Z. Jia}
\affiliation{Quantum Optoelectronics Laboratory, School of Physical Science and Technology, Southwest Jiaotong University, Chengdu 610031, P. R. China}
\author{\ Yong Li}
\affiliation{Beijing Computational Science Research Center, Beijing 100084, P. R. China}
\author{\ Jin-Hui Wu}
\email{jhwu@jlu.edu.cn}
\affiliation{College of Physics, Jilin University, Changchun 130012, P. R. China}

\date{\today }


\begin{abstract}
We study the optomechanically induced amplification and perfect transparency in a double-cavity optomechanical system. We find if two control lasers with appropriate amplitudes and detunings are applied to drive the system, the
phenomenon of optomechanically induced amplification for a probe laser can occur. In addition, perfect optomechanically
induced transparency phenomenon, which is robust to mechanical dissipation, can be realized by the same type of drive. These results are very important for signal amplification, light storage, fast light and slow light in the quantum information processes.

\end{abstract}
\maketitle

\section{Introduction}

Cavity optomechanics, exploring the interaction between light fields and
mechanical motions, has attracted a lot of attention in the past few years for
its potential application in the ultrasensitive detection of tiny mass, force,
and displacement \cite{A1,A2,A3,A4}. One standard and simplest optomechanical
setup is a Fabry-Perot cavity with one end mirror being a micro- or
nano-mechanical vibrating object \cite{B1,B2,B3}. Other various optomechanical
experimental system are designed and investigated such as silica toroidal
optical microresonators \cite{C1,C2,C3}, photonic crystal cavities \cite{D1,D2},
micromechanical membranes \cite{E1,E2}, typical optomechanical cavities confining cold
atoms \cite{f1,f2}, superconducting circuits \cite{g1,g2}, and so on.

Typically, when driving an otomechanical cavity by a red-detuned laser,  the
mechanical oscillator can be cooled to its quantum
ground-state \cite{h1,h2,h3}. Moreover,
in this red-detuned regime, some well-known phenomena in atomic ensemble can find their analogy in optomechanical system. Specifically, under a strong driving, normal mode splitting \cite{j1,j2}(called Autler-Townes effects in atomic physics) can be observed. On the contrary, for a relatively weak driving (much less than the cavity dissipation rate),  an electromagnetically induced transparency like phenomenon, called optomechanically induced transparency\cite{L1,L2,L3},  has been theoretically predict and experimentally verified. This phenomenon can be used to slow down and even stop light signals
\cite{M1,M2} in the long-lived mechanical vibrations.
On the other hand,
when a driving laser applied on the mechanical blue sideband, the
mechanical element of an optomecanical system can be heated, leading to phonon lasing \cite{i1,i2,i3}and probe amplification \cite{p1,p2,N1,N2}.

In our previous work, we have investigated coherent perfect transmission and absorption in a double-cavity optomechanical system driven by two pump fields on red mechanical sideband \cite{q1}. While in this paper, we study the
optomechanically induced amplification and perfect transparency in the same system driven under a different type of drive. We find that if driving the double-cavity
optomechanical system by a red sideband laser from one side and a blue sideband one from the other side and appropriately manipulating the amplitudes of them,  optomechanically induced amplification phenomenon can occur for a nearly resonant weak signal
field (probe field).  In addition, by adjusting the control fields, an interesting  perfect
optomechanically induced transparency (with transmission coefficient rigorously equal to 1) can be realized under the same type of drive. When this perfect transmission occur, quantum coherence process due to the double-driving can totally suppress the decoherence due to the dissipation of mechanical resonator. This double-driving device could be used to
realize signal quantum amplifier, quantum switch, quantum memory and so on.

The rest of this paper is organized as follows. In Section II, we introduce
the double-cavity optomechanical model, obtain the equations of motion for the
mechanical resonator and the two cavity modes, and solve them and obtain the
output fields. In Section III, we show how to realize perfect optomechanically
induced transparency even though with big mechanical decay rate $\gamma_{m}$.
In Section IV, we show how to realize optomechanically induced amplification about the
weak signal field (probe field), meanwhile, the system holds below the phonon
lasing threshold. And the conclusions are given in the Section V.

\section{Model and Equations}

\begin{figure}[ptbh]
\includegraphics[width=0.45\textwidth]{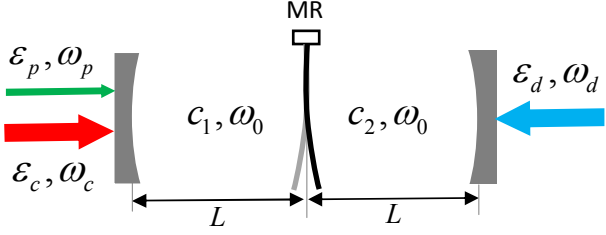}\caption{(Color online) A
double-cavity optomechanical system with a mechanical resonator (MR) inserted
between two fixed mirrors. The two cavities have identical cavity lengths $L$
and mode frequencies $\omega_{0}$ in the absence of radiation pressure.
Coupling field and driving field with frequencies $\omega_{c}, \omega_{d}$ and
amplitudes $\varepsilon_{c}, \varepsilon_{d}$ respectively, act upon opposite
sides of the double-cavity system. The probe field with frequency $\omega_{p}$
and amplitude $\varepsilon_{p}$ is injected into the left optical cavity.}%
\label{Fig1}%
\end{figure}

We consider a double-cavity hybrid system with one mechanical resonator (MR)
of perfect reflection inserted between two fixed mirrors of partial
transmission (see Fig. 1). The MR has an eigen frequency $\omega_{m}$ and a
decay rate $\gamma_{m}$ and thus exhibits a mechanical quality factor
$Q=\omega_{m}/\gamma_{m}$. Two identical optical cavities of lengths $L$ and
frequencies $\omega_{0}$ are got when the MR is at its equilibrium position in
the absence of external excitation. We describe the two optical modes,
respectively, by annihilation (creation) operators $c_{1}$ ($c_{1}^{\dagger}$)
and $c_{2}$ ($c_{2}^{\dagger}$) while the only mechanical mode by $b$
($b^{\dagger}$). These annihilation and creation operators are restricted by
the commutation relation $[c_{i},c_{i}^{\dagger}]=1$ ($i=1,2$) , $[c_{1}%
,c_{2}]=0$, and $[b,b^{\dagger}]=1$. Two coupling fields are used to drive the
double-cavity system from either left or right fixed mirrors with their
amplitudes denoted by $\varepsilon_{c}=\sqrt{2\kappa\wp_{c}/(\hbar\omega_{c}%
)}$ and $\varepsilon_{d}=\sqrt{2\kappa\wp_{d}/(\hbar\omega_{d})}$ and one
probe field is injected into the left optical cavity with amplitude denoted by
$\varepsilon_{p}=\sqrt{2\kappa\wp_{p}/(\hbar\omega_{p})}$. Here $\wp_{c}$,
$\wp_{d}$ and $\wp_{p}$, are relevant field powers,\ $\kappa$ is the common
decay rate of both cavity modes, and $\omega_{c}$, $\omega_{d}$, and
$\omega_{p}$, are relevant field frequencies. Then the total Hamiltonian in
the rotating-wave frame of frequency $\omega_{c}+\omega_{d}$ can be written
as
\begin{align}
H  &  =\hbar\Delta_{c}c_{1}^{\dagger}c_{1}+\hbar\Delta_{d}c_{2}^{\dagger}%
c_{2}+\hbar g_{0}(c_{2}^{\dagger}c_{2}-c_{1}^{\dagger}c_{1})(b^{\dagger
}+b)\label{Eq1}\\
&  +\hbar\omega_{m}b^{\dagger}b+i\hbar\varepsilon_{c}(c_{1}^{\dagger}%
-c_{1})+i\hbar\varepsilon_{d}(c_{2}^{\dagger}-c_{2})\nonumber\\
&  +i\hbar(c_{1}^{\dagger}\varepsilon_{p}e^{-i\delta t}-c_{1}\varepsilon
_{p}^{\ast}e^{i\delta t})\nonumber
\end{align}
with $\Delta_{c}=\omega_{0}-\omega_{c}$ ($\Delta_{d}=\omega_{0}-\omega_{d}$)
being the detuning between cavity modes and coupling field (driving field),
$\delta=\omega_{p}-\omega_{c}$ being the detuning between the probe field and
the coupling field, and $g_{0}=\frac{\omega_{0}}{L}\sqrt{\frac{\hbar}%
{2m\omega_{m}}}$ being the hybrid coupling constant between mechanical and
optical modes.

The dynamics of the system is described by the quantum Langevin equations
for relevant annihilation operators of mechanical and optical modes%
\begin{align}
\dot{b}  &  =-i\omega_{m}b-ig_{0}(c_{2}^{\dagger}c_{2}-c_{1}^{\dagger}%
c_{1})-\frac{\gamma_{m}}{2}b+\sqrt{\gamma_{m}}b_{in},\label{Eq2}\\
\dot{c}_{1}  &  =-[\kappa+i\Delta_{c}-ig_{0}(b^{\dagger}+b)]c_{1}%
+\varepsilon_{c}+\varepsilon_{p}e^{-i\delta t}+\sqrt{2\kappa}c_{1}%
^{in},\nonumber\\
\dot{c}_{2}  &  =-[\kappa+i\Delta_{d}+ig_{0}(b^{\dagger}+b)]c_{2}%
+\varepsilon_{d}+\sqrt{2\kappa}c_{2}^{in}\nonumber
\end{align}
with $b_{in}$ being the thermal noise on the MR with zero mean value,
$c_{1}^{in}$ ($c_{2}^{in}$) is the input quantum vacuum noise from the left
(right) cavity with zero mean value. Because we deal with the mean response of
the system, we do not include these noise terms in the discussion that
follows. In the absence of probe field $\varepsilon_{p}$, Eq.s (2) can be
solved with the factorization assumption $\left\langle bc_{i}\right\rangle
=\left\langle b\right\rangle \left\langle c_{i}\right\rangle $ to generate the
steady-state mean values%
\begin{align}
\langle b\rangle &  =b_{s}=\frac{-ig_{0}(\left\vert c_{2s}\right\vert
^{2}-\left\vert c_{1s}\right\vert ^{2})}{\frac{\gamma_{m}}{2}+i\omega_{m}%
},\label{Eq3}\\
\langle c_{1}\rangle &  =c_{1s}=\frac{\varepsilon_{c}}{\kappa+i\Delta_{1}%
},\nonumber\\
\langle c_{2}\rangle &  =c_{2s}=\frac{\varepsilon_{d}}{\kappa+i\Delta_{2}%
}\nonumber
\end{align}
with $\Delta_{1,2}=\Delta_{c,d}\mp g_{0}(b_{s}+b_{s}^{\ast})$ denoting the
effective detunings between cavity modes and coupling field, driving field
when the membrane oscillator deviates from its equilibrium position. Note in
particular, that $g_{0}\left\vert b_{s}\right\vert $ is typically very small
as compared to $\omega_{m}$ and becomes even exactly zero in the case of
$\left\vert c_{1s}\right\vert =\left\vert c_{2s}\right\vert $ ($\left\vert
\varepsilon_{c}\right\vert =\left\vert \varepsilon_{d}\right\vert $).

In the presence of probe field, however, we can write each operator as the sum
of its mean value and its small fluctuation ($b=b_{s}+\delta b,c_{1}%
=c_{1s}+\delta c_{1},c_{2}=c_{2s}+\delta c_{2}$) to solve Eq. (2) when the
coupling field and the driving field are sufficiently strong. Then keeping
only the linear terms of fluctuation operators and moving into an interaction
picture by introducing $\delta b\rightarrow\delta be^{-i\omega_{m}t}$, $\delta
c_{1}\rightarrow\delta c_{1}e^{-i\Delta_{1}t}$, $\delta c_{2}\rightarrow\delta
c_{2}e^{-i\Delta_{2}t}$, we obtain the linearized quantum Langevin equations%
\begin{align}
\delta\dot{b} &  =-ig_{0}(c_{2s}^{\ast}\delta c_{2}e^{-i(\Delta_{2}-\omega
_{m})t}-c_{1s}^{\ast}\delta c_{1}e^{-i(\Delta_{1}-\omega_{m})t})\label{Eq4}\\
&  -ig_{0}(c_{2s}\delta c_{2}^{\dagger}e^{i(\Delta_{2}+\omega_{m})t}%
-c_{1s}\delta c_{1}^{\dagger}e^{i(\Delta_{1}+\omega_{m})t})-\frac{\gamma_{m}%
}{2}\delta b,\nonumber\\
\delta\dot{c}_{1} &  =-\kappa\delta c_{1}+ig_{0}c_{1s}(\delta be^{-i(\omega
_{m}-\Delta_{1})t}+\delta b^{\dagger}e^{i(\omega_{m}+\Delta_{1})t})\nonumber\\
&  +\varepsilon_{p}e^{-i(\delta-\Delta_{1})t},\nonumber\\
\delta\dot{c}_{2} &  =-\kappa\delta c_{2}-ig_{0}c_{2s}(\delta be^{-i(\omega
_{m}-\Delta_{2})t}+\delta b^{\dagger}e^{i(\omega_{m}+\Delta_{2})t}).\nonumber
\end{align}
If the coupling field drives at the mechanical red sideband while the driving
field drives at the blue sideband ($\Delta_{1}\approx\omega_{m}$, $\Delta
_{2}\approx-\omega_{m}$), the hybrid system is operating in the resolved
sideband regime ($\omega_{m}>>\kappa$), the membrane oscillator has a high
mechanical quality factor ($\omega_{m}>>\gamma_{m}$), and the mechanical
frequency $\omega_{m}$ is much larger than $g_{0}\left\vert c_{1s}\right\vert
$ and $g_{0}\left\vert c_{2s}\right\vert $, Eq.s (4) will be simplified to
\begin{align}
\delta\dot{b} &  =-ig_{0}(c_{2s}\delta c_{2}^{\dagger}-c_{1s}^{\ast}\delta
c_{1})-\frac{\gamma_{m}}{2}\delta b,\label{Eq5}\\
\delta\dot{c}_{1} &  =-\kappa\delta c_{1}+ig_{0}c_{1s}\delta b+\varepsilon
_{L}e^{-ixt},\nonumber\\
\delta\dot{c}_{2} &  =-\kappa\delta c_{2}-ig_{0}c_{2s}\delta b^{\dagger
}\nonumber
\end{align}
with $x=\delta-\omega_{m}$. We can examine the expectation values of small
fluctuations by the following three coupled dynamic equations
\begin{align}
\left\langle \delta\dot{b}\right\rangle  &  =-ig_{0}(c_{2s}\left\langle \delta
c_{2}^{\dagger}\right\rangle -c_{1s}^{\ast}\left\langle \delta c_{1}%
\right\rangle )-\frac{\gamma_{m}}{2}\left\langle \delta b\right\rangle
,\label{Eq6}\\
\left\langle \delta\dot{c}_{1}\right\rangle  &  =-\kappa\left\langle \delta
c_{1}\right\rangle +ig_{0}c_{1s}\left\langle \delta b\right\rangle
+\varepsilon_{p}e^{-ixt},\nonumber\\
\left\langle \delta\dot{c}_{2}\right\rangle  &  =-\kappa\left\langle \delta
c_{2}\right\rangle -ig_{0}c_{2s}\left\langle \delta b^{\dagger}\right\rangle.
\nonumber
\end{align}
We assume  the steady-state solutions of above equations have form: $\langle\delta s\rangle=\delta s_{+}e^{-ixt}+\delta s_{-}e^{ixt}$
with $s=b,c_{1},c_{2}$. Then it is straightforward to obtain the following
results
\begin{align}
\delta b_{+} &  =\frac{iG\varepsilon_{p}}{(\kappa-ix)(\frac{\gamma_{m}}%
{2}-ix)+G^{2}(1-n^{2})},\\
\delta c_{1+} &  =\frac{\varepsilon_{p}[-n^{2}G^{2}+(\kappa-ix)(\frac
{\gamma_{m}}{2}-ix)]}{(\kappa-ix)^{2}(\frac{\gamma_{m}}{2}-ix)+G^{2}%
(1-n^{2})(\kappa-ix)},\nonumber\\
\delta c_{2-} &  =\frac{-nG^{2}\varepsilon_{p}}{(\kappa+ix)^{2}(\frac
{\gamma_{m}}{2}+ix)+G^{2}(1-n^{2})(\kappa+ix)},\nonumber
\end{align}
where $G=g_{0}c_{1s}$ is the effective optomechanical coupling
rate and $\left\vert c_{2s}/c_{1s}\right\vert ^{2}=n^{2}$ is the photon number
ratio of two cavity modes. In deriving Eqs. (7), we have also assumed that
$c_{1s,2s}$ is real-valued without loss of generality.

Based on Eqs. (7), we can further determine the left-hand output field
$\varepsilon_{outL}$ and the right-hand output field $\varepsilon_{outR}$
through the following input-output relation \cite{Walls} \bigskip%
\begin{align}
\varepsilon_{outL}  &  =2\kappa\langle\delta c_{1}\rangle-\varepsilon
_{p}e^{-ixt}\label{Eq8}\\
\varepsilon_{outR}  &  =2\kappa\langle\delta c_{2}\rangle,\nonumber
\end{align}

where the oscillating terms can be removed if we set $\varepsilon
_{outL}=\varepsilon_{outL+}e^{-ixt}+\varepsilon_{outL-}e^{ixt}$ and
$\varepsilon_{outR}=\varepsilon_{outR+}e^{-ixt}+\varepsilon_{outR-}e^{ixt}$.
Note that the output components $\varepsilon_{outL+}$ and $\varepsilon
_{outR-}$ have the same frequency $\omega_{p}$\ as the input probe
fields $\varepsilon_{p}$, while the output components $\varepsilon_{outL-}$ and
$\varepsilon_{outR+}$ are generated at frequencies $2\omega_{c}-\omega_{p}$ and $2\omega_{d}-\omega_{p}$, respectively, in a nonlinear wave-mixing process of optomechanical
interaction. Then with Eqs. (8) we can obtain
\begin{align}
\varepsilon_{outL+}  &  =2\kappa\delta c_{1+}-\varepsilon_{p},\label{Eq9}\\
\varepsilon_{outR-}  &  =2\kappa\delta c_{2-}\nonumber
\end{align}
oscillating at frequency $\omega_{p}$ of our special interest.

In this paper, we discuss the perfect optomechanically induced amplification and transparency
under the realistic parameters in a
optomechanical experiment \cite{j2}. That is, $L=25$ mm, $m=145$
ng, $\kappa=2\pi\times215$ kHz, $\omega_{m}=2\pi\times947$ kHz, and
$\gamma_{m}=2\pi\times141$ Hz. In addition, the laser wavelength is $\lambda$
$=$ $2\pi c/\omega_{c}=1064$ nm and the mechanical quality factor is $Q=$
$\omega_{m}/\gamma_{m}=6700$.

\section{Perfect optomechanically induced transparency}

Now we study the perfect optomechanically induced transparency for the probe
field. The quadrature of the optical components with
frequency $\omega_{p}$ in the output field can be defined as  $\varepsilon_{T}=2\kappa\delta c_{1+}/\varepsilon_{p}$
\cite{L2} . Specifically, it can be written as

\begin{align}
\varepsilon_{T} =\frac{2\kappa[-n^{2}G^{2}+(\kappa-ix)(\frac{\gamma_{m}}%
{2}-ix)]}{(\kappa-ix)^{2}(\frac{\gamma_{m}}{2}-ix)+G^{2}(1-n^{2})(\kappa-ix)},
\end{align}
whose real and imaginary part ($Re[\varepsilon_{T}]$ and $Im[\varepsilon_{T}]$)
represent the absorptive and dispersive behavior of the optomechanical system,
respectively. It is well-known that in a standard optomechanical system with single optical cavity, the optomechanically induced transparency
dip is not perfect as the decay $\gamma_{m}$ of the mechanical resonator is
not zero. However, we can see from Eq.(10) that, in the double-cavity optomechanical system studied here, if setting the ratio $n=\sqrt{\gamma_{m}%
\kappa/2G^{2}}$, the optomechanically induced transparency dip will be perfect
even though remarkale mechanical decay $\gamma_{m}$ exists.

\begin{figure}[ptbh]
\centering \includegraphics[width=0.45\textwidth]{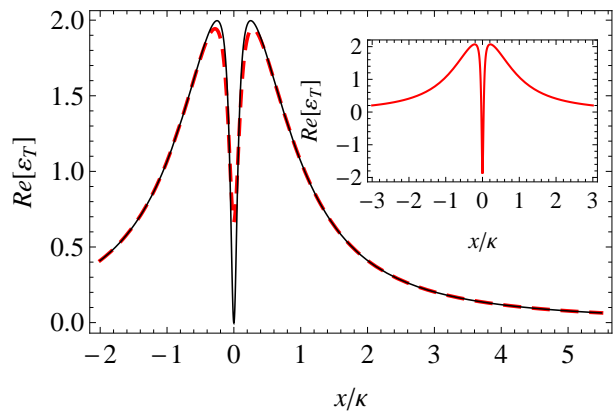}\caption{The real
part of $\varepsilon_{T}$ vs. the normalized frequency detuning $x/\kappa$:
$n=0$ (red-dashed) and $n=0.7$ (black-solid) with $\gamma_{m}=2\pi\times14.1$
kHz and $\wp_{c}=1mW$. In the inset: $n=0.7$, $\gamma_{m}=2\pi\times141$ Hz
and $\wp_{c}=1mW$.}%
\label{Fig2}%
\end{figure}

\begin{figure}[ptbh]
\centering \includegraphics[width=0.45\textwidth]{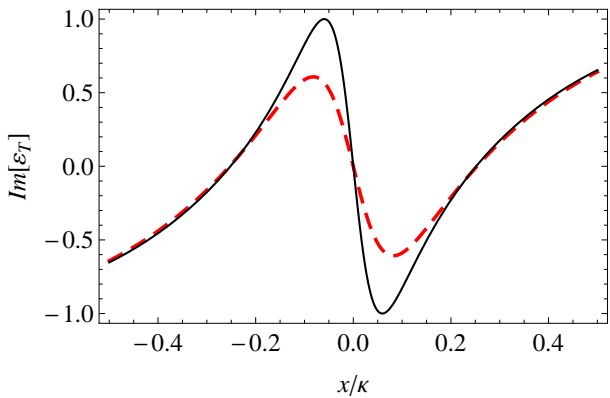}\caption{The imaginary
part of $\varepsilon_{T}$ vs. the normalized frequency detuning $x/\kappa$:
$n=0$ (red-dashed) and $n=0.7$ (black-solid) with $\gamma_{m}=2\pi\times14.1$
kHz and $\wp_{c}=1mW$.}%
\label{Fig3}%
\end{figure}

To see this clearly, in Fig. 2, we plot the $Re[\varepsilon_{T}]$ versus the
normalized frequency $x/\kappa$ with $\gamma_{m}=2\pi\times14.1$ kHz and
$\wp_{c}=1mW$ for different $n$. We can see form Fig. 2 that when $n=0$ (i.e. the usual optomechanically induced transparency case), the
optomechanically induced transparency dip will become shallow with a large
mechanical decay $\gamma_{m}$  (red-dashed). However, when an additional blue-sideband driving field satisfying the condition $n=\sqrt{\gamma_{m}\kappa/2G^{2}}\approx0.7$ applied,  the transparency dip will become
perfect, exhibiting totally transmission of probe laser (black-solid). Physically,
it means that the dissipative energy through the decay
$\gamma_{m}$ of the mechanical resonator can be compensated by applying the
right-hand driving field with amplitude $\varepsilon_{d}=\varepsilon_{c}%
\sqrt{\gamma_{m}\kappa/2G^{2}}$ and the blue mechanical sideband frequency.
When $\omega_{p}\approx\omega_{0}$, $n=\sqrt{\gamma_{m}\kappa/2G^{2}}$ and the
beat frequency $\omega_{p}-\omega_{c}=\omega_{m} (x=0)$, thus, the MR is
driven by a force oscillating at its eigenfrequency $\omega_{m}$ and the
resonator starts to oscillate coherently. This motion will generate photons
with frequency $\omega_{p}$ that interfere destructively with the probe beam,
leading to a optomechanically induced transparency dip.

In Fig. 3, we plot the the dispersion curve $Im[\varepsilon_{T}]$ versus the
normalized frequency $x/\kappa$ with $\gamma_{m}=2\pi\times14.1$ kHz and
$\wp_{c}=1mW$ for different $n$. Clearly, the curve with $n=0.7$ (black-solid) is much steeper than the one with $n=0$ (red-dashed) in the vicinity of $x=0$. It means that we can easily control the dispersive behavior of the optomechanical system by applying  the blue-detuned driving field with amplitude $\varepsilon_{d}=n\varepsilon_{c}$, which can possibly be used to control slow light in optomechanical systems.

\section{optomechanically induced amplification}

In this section, we study the optomechanically induced amplification in this double-cavity optomechanical system. If the ratio $n>\sqrt{\gamma_{m}\kappa/2G^{2}}$, we find the $Re[\varepsilon
_{T}]$ will become negative in the vicinity of $x=0$ (see the inset in Fig.
2). It means that optomechanically induced gain (amplification) can be
realized in this double-cavity system by applying a blue-detuned driving field to the right-side cavity
with amplitude $\varepsilon_{d}=n\varepsilon_{c}$.  Note that when the system works under the condition
$x=0$, $n=\sqrt
{1+\frac{\gamma_{m}\kappa}{2G^{2}}}$, $Re[\varepsilon
_{T}]$ will be divergent. In addition, the system will work into the
parametric instability regime as $n\gtrsim1$ when the input power $\wp_{c}%
=1$mW, and so we limit ourselves to the case where $n\leq1$.

\begin{figure}[ptbh]
\centering \includegraphics[width=0.45\textwidth]{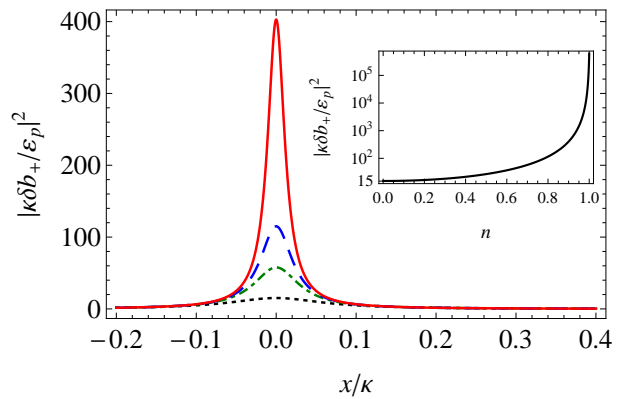}\caption{The
normalized mechanical oscillation $|\kappa\delta b_{+}/\varepsilon_{p}|^{2}$
vs. the normalized frequency detuning $x/\kappa$: $n=0$ (black-dotted),
$n=0.7$ (green-dotted-dashed), $n=0.8$ (blue-dashed), and $n=0.9$ (red-solid)
with $\wp_{cL}=1mW$. In the inset, we plot the normalized mechanical
oscillation $|\kappa\delta b_{+}/\varepsilon_{p}|^{2}$ vs. the ratio $n$.}%
\label{Fig4}%
\end{figure}

In Fig. 4, we plot the mechanical oscillation $|\kappa\delta b_{+}%
/\varepsilon_{p}|^{2}$ (normalized to probe field $\varepsilon_{p}$) versus
the normalized frequency $x/\kappa$ for different $n$. In the inset we plot
the $|\kappa\delta b_{+}/\varepsilon_{p}|^{2}$ as a function of $n$ for $x=0$.
It can be seen clearly from Fig. 4 that the mechanical oscillation peak value
locates at $x=0$, and increases with $n$ [$n=0$
(black-dotted), $n=0.7$ (green-dotted-dashed), $n=0.8$ (blue-dashed), $n=0.9$
(red-solid)]. And when $n$ increases up to 1, the mechanical oscillation peak
value will increase approximately to $6.1\times10^{5}$ (see the inset in Fig.
4). It means that the optomechanical effect will become stronger for bigger
$n$ (less than or equal to 1) when $\omega_{p}-\omega_{c}=\omega_{m} (x=0)$
and $\omega_{d}-\omega_{0}=\omega_{m}$. The reason for this is that, the
blue-mechanical sideband (heating sideband) of right-hand cavity generating
much phonons which will be absorbed by the Anti-Stokes processes in left-hand
cavity for the red-mechanical sideband (cooling sideband). Then, the
optomechanical effect of the double-cavity system is resonantly enhanced.

\begin{figure}[ptbh]
\centering \includegraphics[width=0.45\textwidth]{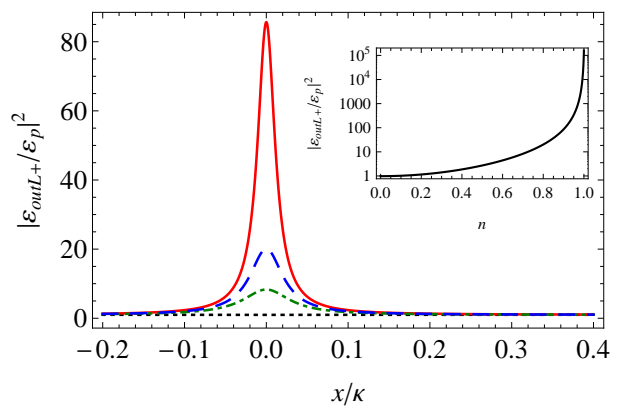}\caption{The
normalized left-hand output energy $|\varepsilon_{outL+}/\varepsilon_{p}|^{2}$
vs. the normalized frequency detuning $x/\kappa$: $n=0$ (black-dotted),
$n=0.7$ (green-dotted-dashed), $n=0.8$ (blue-dashed), and $n=0.9$ (red-solid)
with $\wp_{cL}=1mW$. In the inset, we plot the normalized output energy
$|\varepsilon_{outL+}/\varepsilon_{p}|^{2}$ vs. the ratio $n$.}%
\label{Fig5}%
\end{figure}

\begin{figure}[ptbh]
\centering \includegraphics[width=0.45\textwidth]{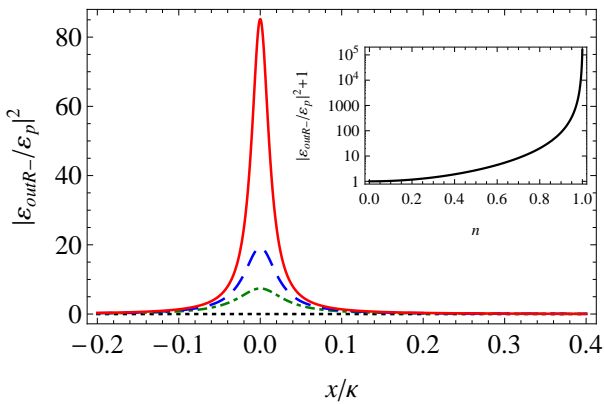}\caption{The
normalized right-hand output energy $|\varepsilon_{outR-}/\varepsilon_{p}%
|^{2}$ vs. the normalized frequency detuning $x/\kappa$: the parameters are
the same as in Fig. 4. In the inset, we plot the normalized output energy
$|\varepsilon_{outR-}/\varepsilon_{p}|^{2}+1$ vs. the ratio $n$.}%
\label{Fig6}%
\end{figure}

In Fig. 5-6, we plot the output power $|\varepsilon_{outL+}/\varepsilon
_{p}|^{2}$ and $|\varepsilon_{outR-}/\varepsilon_{p}|^{2}$ normalized to the
input probe field $\varepsilon_{p}$ respectively, versus the normalized
frequency $x/\kappa$ for different $n$. It can be seen clearly from Fig. 5-6
that the output energies $|\varepsilon_{outL+}/\varepsilon_{p}|^{2}$ and
$|\varepsilon_{outR-}/\varepsilon_{p}|^{2}$ get the maximum value at $x=0$ for
a certain value $n$. When $x=0$, the output normalized energies $|\varepsilon
_{outL+}/\varepsilon_{p}|^{2}$ and $|\varepsilon_{outR-}/\varepsilon_{p}|^{2}$
will increase with $n$ which is similar to the mechanical
oscillation $|\kappa\delta b_{+}/\varepsilon_{p}|^{2}$. This is because that
when $x=0$, the optomechanical effect will be strongest for a certain value
$n$ as discussed above. The curves of the output normalized energies
$|\varepsilon_{outL+}/\varepsilon_{p}|^{2}$ and $|\varepsilon_{outR-}%
/\varepsilon_{p}|^{2}$ almost have the same line shape, except that the output
normalized energy $|\varepsilon_{outL+}/\varepsilon_{p}|^{2}$ starts from 1
with the increase of $n$ for $x=0$ while the output normalized energy
$|\varepsilon_{outR-}/\varepsilon_{p}|^{2}$ starts from 0 (see the insets in
Fig. 5-6). This shows that the double-cavity optomechanical system will be reduced to
the standard one-cavity optomechanical model ($|\varepsilon_{outR-}%
/\varepsilon_{p}|^{2}=0$) when $n=0$.  When $n$ increases up to 1, the
output normalized energies $|\varepsilon_{outL+}/\varepsilon_{p}|^{2}$ and
$|\varepsilon_{outR-}/\varepsilon_{p}|^{2}$ will increase approximately to
$1.6\times10^{5}$ (see the insets in Fig. 5-6). The reason for this is that,
the existing of blue-detuned driving field with $\omega_{d}-\omega_{0}=\omega_{m}%
$ will coherently enhance the oscillation of the MR (see Fig. 4),
leading to optomechanically induced amplification. Thus, we can realize the
optomechanically induced amplification for a resonantly injected probe in the double-cavity optomechanical system by appropriately
adjusting the ratio $n$ of the two strong field amplitudes
$\varepsilon_{c,d}$.

\section{Conclusions}

In summary, we have studied in theory a double-cavity optomechanical system
driven by a red sideband laser from one side and a blue sideband one from the other side. Our analytical and numerical results show that if adjusting the amplitude-ratio of the two driving fields $n>\sqrt{\gamma_{m}\kappa/2G^{2}}$ , the optomechanically induced
amplification for a resonantly incident probe (i.e., $\omega_{p}-\omega_{c}-\omega_{m}=0$) can be realized in this system. Typically, remarkable amplification can be obtained when $n\sim1$.
The reason for this is that, the Stokes processes in the blue-sideband driven cavity can generate
 phonons in the mechanical elements, and these phonons will be further absorbed by the
Anti-Stokes processes in the red-sideband driven cavity. As a result, the optomechanical effect of the double-cavity
system is resonantly enhanced. In addition, the perfect optomechanically
induced transparency can be realized if we set the ratio $n=\sqrt
{\gamma_{m}\kappa/2G^{2}}$. Different from usual optomechanical induced transparency, this phenomenon is robust to mechanical dissipation, namely, the perfect transparency window can preserve even if the mechanical resonator has a relatively large decay rate $\gamma_{m}$. We expect that our study can be used to realize
signal quantum amplifier and light storage in the quantum information processes.

\begin{acknowledgments}
This work is supported by the National Natural Science Foundation of China (61378094). W. Z. Jia is supported by the National Natural Science Foundation of China under Grant No. 11347001 and the Fundamental Research Funds for the Central Universities 2682014RC21.
\end{acknowledgments}

\end{document}